\documentclass[preprint,aip,jap,amsfonts,amssymb,amsmath,floatfix,showpacs]{revtex4-1}
\usepackage[]{graphicx}% Include figure files
\usepackage{epstopdf}
\usepackage{bm}
\usepackage{hyperref}
\usepackage{framed}
\usepackage{wasysym}
\usepackage{dcolumn}% Align table columns on decimal point
\usepackage{bm}% bold math
\newcommand{\eg}{e.g.,}
\newcommand{\ie}{i.e.,}
\newcommand{\etal}{\textit{et al.}}
\begin{document}
\title{How pump-probe differential reflectivity at negative delay yields the perturbed free-induction-decay: Theory of the experiment and its verification}
\author{Richarj Mondal}
\affiliation{Indian Institute of Science Education and Research Kolkata, Mohanpur, Nadia 741246, West Bengal, India}
\author{Basabendra Roy}
\affiliation{Indian Institute of Science Education and Research Kolkata, Mohanpur, Nadia 741246, West Bengal, India}
\author{Bipul Pal}\email{bipul@iiserkol.ac.in}
\affiliation{Indian Institute of Science Education and Research Kolkata, Mohanpur, Nadia 741246, West Bengal, India}
\author{Bhavtosh Bansal} \email{bhavtosh@iiserkol.ac.in}
\affiliation{Indian Institute of Science Education and Research Kolkata, Mohanpur, Nadia 741246, West Bengal, India}
\date{\today}
\begin{abstract}
While time-resolved pump-probe differential reflectivity and transmitivity measurements are routinely used to monitor the population relaxation dynamics on the subpicosecond time scale, it is also known that the signal in the negative delay can yield direct signatures of the perturbed-free-induction-decay of polarization. Yet this technique, especially in reflection geometry, has never been popular because the experiment is conceptually not very intuitive. Coherent dynamics is therefore usually studied using the more complex four-wave-mixing experiments. Here we derive from first principles the simplest possible but mathematically complete framework for the negative delay signal in both the time and the spectral domains. The calculation involving the optical Bloch equations to describe the induced polarization and the Ewald-Oseen idea to calculate the reflected signal as a consequence of the free oscillations of perturbed dipoles, also explicitly includes the process of lock-in detection of a double-chopped signal after it has passed through a monochromator. The theory is compared with experiments on high quality GaAs quantum well sample. The dephasing time inferred experimentally at 4 K compares remarkably well with the inverse of the absorption linewidth of the continuous-wave photoluminescence excitation spectrum. Spectrally resolved signal at negative delay calculated from our theoretical expression nicely reproduces the coherent spectral oscillations, although exact fitting of the experimental spectra with the theoretical expression is difficult. This is  on account of additional resonances present in the sample corresponding to lower energy bound states.
\end{abstract}

\maketitle
\date{\today}
\section{\label{sec:level1}Introduction}
The dynamics of excitons in the coherent regime ($\lesssim 10$~ps) after resonant excitation  by ultrashort laser pulses has been an active field of research for over four decades now.~\cite{shah_book, Tsen, Meier-Thomas-Koch, Chemla, Schultheis, wang, Honold, pal4, Koch, Vinattieri, Felix, Guenther, pal1, pal2, Fluegel, Bartels, Neukirch, Sokoloff, Lindberg1, Lindberg2}  These studies give fundamental information about the scattering processes which lead to dephasing. Exciton dephasing dynamics has been studied in the time domain, mainly using nonlinear wave-mixing experiments,~\cite{shah_book, Tsen, Meier-Thomas-Koch, Chemla} \textit{viz.} four-wave-mixing, photon echo and phase conjugation. Four-wave-mixing  (FWM)~\cite{shah_book, Tsen, Meier-Thomas-Koch, Chemla, Schultheis, wang, Honold, pal4, Erland, Shacklette, pal3, pal5,yajima} relies on the temporal blurring of the optical grating formed by the standing wave pattern generated within a sample on account of the interfering polarizations induced by two incident laser beams. Wave-mixing experiments are usually analyzed by solving the optical Bloch equations for the appropriate nonlinear polarization (third order for FWM) having the required wave vector which decides the signal propagation direction. Time-integrated FWM signal at positive delay gives information about exciton polarization dephasing.

Pump-probe differential transmitivity (PPDT) or pump-probe differential reflectivity (PPDR) experiments are much simpler than these wave-mixing techniques and have been used for a variety of semiconductor samples to investigate many interesting phenomena such as exciton ionization, broadening and shift of excitonic transition, etc.~\cite{Knox,Wegener,Nisoli,Hulin,Tai, Cruz} In PPDT and PPDR experiments, like DFWM experiments,~\cite{fnote1} ultrashort pulses from two laser-beams having wave vectors $\vec{k}_{pm}$ (the pump beam) and $\vec{k}_{pr}$ (the probe beam) are cofocused on to the sample. However, unlike DFWM experiments, the PPDT (PPDR) signal arising from the third order nonlinear optical effect is detected  along the probe transmission (reflection) direction, where a linear optical signal is also present. As this facilitates easy identification of the PPDT (PPDR) signal direction, setting up PPDT and PPDR experiments becomes much easier compared to the nonlinear wave-mixing experiments. The pump-probe technique has been conventionally used to study incoherent processes such as energy relaxation and recombination dynamics in a large number of materials under conditions of positive delay (pump preceding the probe).~\cite{shank, doany1, doany2, gupta, kadow, hanson} While the theory of susceptibility at negative delays has long been developed~\cite{Meier-Thomas-Koch,Cruz} and there are many reports of measurements of the perturbed-free-induction-decay using negative delay pump-probe measurements~\cite{Guenther, pal1, Fluegel, Bartels, Neukirch, Sokoloff, Lindberg1, Lindberg2} (mostly in transmission geometry),~\cite{Fluegel, Bartels, Neukirch, Sokoloff, Lindberg1, Lindberg2} its overall use has been limited. This is primarily because the interpretation of the measured signal is not quite transparent and has sometimes been  controversial.~\cite{Joffre} It is hard to intuitively understand why the spectrally-resolved, time-integrated PPDT (PPDR) signal is non-zero at negative delays, how it carries information of polarization dephasing within the sample, or why the time- and spectrally-integrated PPDT (PPDR) signal is zero at negative delays.~\cite{Guenther, Joffre} While existing theories adequately deal with the third order polarization or susceptibility responsible for the PPDT (PPDR) signal, a complete calculation deriving the perturbatively generated higher order reflected beam and detailing the role of the double-modulation of the pump and probe beams, the spectral filtering by the monochromator, the time-integrated measurement by the slow detector and the phase-sensitive detection of the signal by the lock-in amplifier (accounting for all the factors of $\pi$, etc.) would be, we believe, useful.

In this paper we attempt to develop such a theoretical framework for the pump-probe differential reflectivity measurements at negative delay. The paper is organized as follows. First we present a description of a typical PPDR experimental setup to identify the important components of the experimental scheme. Then we develop our theoretical model of PPDR signal taking account of each of the important experimental components. Finally, we compare our theoretical model with the results of PPDR experiments on high quality GaAs/AlGaAs multi-quantum well sample at low temperature. The theory nicely reproduced the coherent spectral oscillations observed in low-temperature PPDR spectra at negative delays. The dephasing time estimated from the perturbed-free-induction-decay in the time-integrated PPDR signal measured at exciton peak at negative delay matches well with the inverse linewidth of the absorption spectra which is estimated from the photoluminescence excitation (PLE) spectroscopy.~\cite{Pelant-Valenta} This provides a strong evidence to the credibility of the pump-probe method.

\section{\label{sec:level2}Experimental}
The schematic of a basic experimental setup for spectrally-resolved time-integrated PPDR measurements, such as the one used for measurements which led to Figs.~3 and 5 below, is shown in Fig.~1. A beam containing train of $\sim 100$~fs pulses at a repetition rate of 80~MHz from a mode-locked Ti:Sapphire laser is split by a beam-splitter into a weak probe-beam and a strong pump-beam. The intensities of the pump and probe beams can be independently controlled by the two variable neutral density filters in the path of the pump and probe beams, respectively. The pulses in the pump beam are controllably delayed with respect to the corresponding pulses in the probe beam by a retroreflector-on-a-linear-translation-stage arrangement. The pump and probe beams are modulated at different frequencies, $f_{pm}=760$~Hz and $f_{pr}=912$~Hz respectively, by a mechanical chopper having square-wave-like chopping profile with a 50{\%} duty cycle. The pump and probe beams are co-focused on to the sample, with each having a spot size of $\sim 50$~$\mu$m. The sample is mounted on to the cold-finger of a closed-cycle helium cryostat fitted with a quartz optical window. This allows for the control of the sample temperature between 4--300~K.  The pump and probe beams are incident on the sample at an angle of $15^0$ and $10^0$ respectively, from the normal to the sample surface and are denoted by the wave vectors $\vec{k}_{pm}$ and $\vec{k}_{pr}$, respectively. Both the beams are linearly polarized and the relative polarization of the two beams are set to be orthogonal to each other by a half-wave plate placed in the path of the pump beam. The reflected pump beam is blocked by a beam-blocker. The reflected probe beam is spectrally dispersed by a monochromator with a narrow spectral band-pass of $\sim 0.1$~nm.  A small aperture is placed in front of the entrance slit of the monochromator to reduce the stray light entering the monochromator. The signal passing through the narrow band-pass of the monochromator is measured with a photomultiplier tube (PMT) detector. The signal from the PMT is fed to a lock-in amplifier which is locked to the sum frequency $f_{pm}+f_{pr}=1.67 $~kHz.~\cite{sandip_double-modulataion} Lock-in time-constant is set to be 300~ms. Locking to the sum (or difference) frequency allows for a direct measurement of the PPDR signal which is the change in the probe reflectivity due to the pump beam. The monochromator, the delay stage, and the lock-in amplifier are synchronously controlled by a computer. One may  measure either the PPDR spectrum at a given delay or measure the delay dependence of the PPDR signal at a given wavelength.
%%%%%%%%%%%%%%%%%%%%%%%%%%%%%%%%%%%%%%%%
\begin{figure}[htb]
    \includegraphics[clip,width=10.0cm]{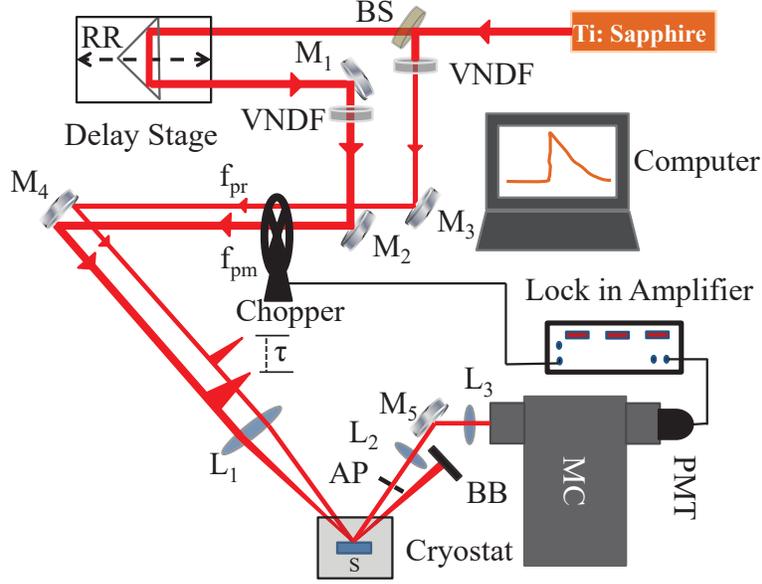}
    \caption{Schematic of the pump-probe differential reflectivity experimental setup. Symbols: BS~=~beam splitter, M$_{1}$--M$_{5}=$~mirrors, L$_{1}$--L$_{4}=$~lenses, VNDF~=~variable neutral density filter, RR~=~retroreflector, $f_{pm}$ and $f_{pr}=$~pump and probe chopping frequencies, AP~=~ aperture, BB~=~beam-blocker, MC~=~monochromator, S~=~sample, and PMT~=~photomultiplier tube detector.}\label{setup}
\end{figure}
%%%%%%%%%%%%%%%%%%%%%%%%%%%%%%%%%%%%%%%%%

\section{\label{sec:level3}Theoretical Calculations}
Let us now model the experimental setup described above in precise mathematical terms, while still keeping the treatment analytically tractable and physically transparent, ensuring that we have closed form expressions in the end. The important steps in the calculation would  be (i) to write down the electric fields due to the pump and the probe pulse trains modulated by the square-wave chopping functions imparted by the optical chopper, (ii) to calculate the relevant components of polarization (having the required wavevector combination) generated in the sample due to the pump and the probe electric fields, (iii) to calculate the radiated electric field in the probe-reflection direction due to the generated polarization within the sample, and (iv) to calculate the detected light intensity due the radiated electric filed in the probe-reflection direction, taking account of the effect of the spectral filtering by the monochromator, time integration of the detected light intensity by the slow detector that averages over many pulses, and phase-sensitive detection by the lock-in amplifier which is locked to the sum frequency of the pump and probe chopping.

Let us start with a form for the electric field from the pulsed Ti:sapphire laser. To keep the treatment analytically tractable, we approximate the $\sim 100$ femtosecond pulses as periodic delta function pulses with a time separation  $\Delta$ between consecutive pulses. The train of pulses in the pump and the probe beams arrive at the sample with a relative delay $\tau$. In this calculation we are interested in negative delays where the probe pulse precedes the corresponding pump pulse. Thus, the electric field due to the probe and pump pulses are written, respectively,  as
\begin{equation}
\vec{E}^0_{pr}(\vec{r},t)=\sum_{m}\{\vec{\zeta}_{pr}\delta(t-m\Delta)\exp[i(\vec{k}_{pr}\cdot\vec{r}-\omega_{L} t)]+\text{c.c.}\},
\end{equation}
and
\begin{equation}
\vec{E}^0_{pm}(\vec{r},t)=\sum_{m}\{\vec{\zeta}_{pm}\delta(t-m\Delta -\tau) \exp[i(\vec{k}_{pm}\cdot\vec{r}-\omega_{L} t)]+\text{c.c.}\}; \qquad  \tau > 0
\end{equation}
where $\vec{\zeta}_{pr}$ ($\vec{\zeta}_{pm}$) is the vector amplitude function which contains the area under the pulse temporal envelope and the unit vector describing the direction of the electric field for the probe (pump) beam,  $\vec{k}_{pr}$ ($\vec{k}_{pm}$) is the wavevector of the probe (pump) beam, $\omega_{L}$  is the frequency of the laser beam which is same for the probe and the pump beams as they both are derived from the same laser, $m$ is an integer, and c.c. stands for the complex conjugate. On account of a strict hierarchy of (almost) non-overlapping time scales, $t_p \sim 100$~fs (pulse width) $<<$  $T_2\sim 1$~ps (polarization dephasing time) $<<$ $\tau_r \sim 100$~ps (population relaxation time) $<<$ $\Delta \sim 10$~ns (pulse-repetition period), each pulse may be treated as a temporal $\delta$-function and the effect of one pulse may be considered to be independent of that of another pulse. Measurements accumulating signals from many pulses only help to improve the signal-to-noise ratio.

If the incident probe and pump beams are modulated by a square waveform imparted by a mechanical chopper having 50{\%} duty cycle with respective chopping frequencies $\Omega_{pr} = 2\pi f_{pr}$ and $\Omega_{pm} = 2\pi f_{pm}$, the Fourier expansions of the respective electric fields of the probe and the pump beams are
\begin{equation} \label{eqpr}
\begin{split}
\vec{E}_{pr}(\vec{r},t,\Omega_{pr})=&\sum_{m}\{\vec{\zeta}_{pr}\delta(t-m\Delta)\exp[i(\vec{k}_{pr}\cdot\vec{r}-\omega_{L} t)]\Bigg[\frac{1}{2}+\frac{2}{\pi}\sum_{n=1,3, 5\ldots}^{\infty} \frac{\sin(n\Omega_{pr}t+\theta_{1})}{n}\Bigg]+\text{c.c}\}\\=&\vec{\xi}^0_{pr}(t,\Omega_{pr})\exp[i(\vec{k}_{pr}\cdot\vec{r}-\omega_{L} t)]+\text{c.c.},
\end{split}
\end{equation}
and
\begin{equation}  \label{eqpm}
\begin{split}
\vec{E}_{pm}(\vec{r},t,\Omega_{pm})=&\sum_{m}\{\vec{\zeta}_{pm}\delta(t-m\Delta-\tau)
\exp[i(\vec{k}_{pm}\cdot\vec{r}-\omega_{L}t)]
\Bigg[\frac{1}{2}+\frac{2}{\pi}\sum_{n=1,3,5\ldots}^{\infty} \frac{\sin(n\Omega_{pm}t+\theta_{2})}{n}\Bigg]+\text{c.c}\}\\=&\vec{\xi}^0_{pm}(t,\Omega_{pm})\exp[i(\vec{k}_{pm}\cdot\vec{r}-\omega_{L} t)]+\text{c.c.}.
\end{split}
\end{equation}
Here $\vec{\xi}^0_{pr}(t,\Omega_{pr})$ and $\vec{\xi}^0_{pm}(t,\Omega_{pm})$ are the amplitudes including the $\delta$-functions and the Fourier series, $\theta_{1}$ is the phase of the probe beam  and $\theta_{2}$ is the phase of the pump beam. The signal-to-noise ratio in the double modulation case is superior as has been reported in the past.~\cite{Frolov,Quitevis} It is worth clarifying that whether one considers the chopper is modulating the electric field or the intensity of the incident beams depends on the phenomenon being probed. In the coherent regime, it is the electric field because the polarization of the dipoles couples to the field,  whereas intensity modulation should be considered for studies in the incoherent regime dealing with carrier population. But note that in either case the Fourier expansion is essentially the same because the squares of equations \eqref{eqpr} and \eqref{eqpm} also yield the same Fourier frequency components (a square-wave train remains as a square-wave train upon squaring its amplitude). It is important to once again emphasize the well-separated time scales appearing in Eqs. \eqref{eqpr} and \eqref{eqpm}. The chopping frequencies are of the order of a few hundred Hz, and $\omega_{L}$ is of the order of $10^{15}$~Hz. Also note that a $\sim 100$~fs laser pulse still contains more than $20$ cycles of the electric field at optical wavelengths.

Under the simplest approximation the sample may be considered to contain an ensemble of independent two-level systems. The interaction with the optical fields of the pump and the probe beams gives rise to macroscopic polarization, which has several components with  different combination of wavevectors $\vec{k}_{pm}$ and $\vec{k}_{pr}$, depending upon the order of interaction considered in the calculation. The time-dependent polarization in turn gives rise to radiating electric fields in different directions depending on the embedded wavevector combinations.  The standard framework for the calculation of the  polarization in the given experimental condition is to solve the optical Bloch equations perturbatively  to the required order.~\cite{shah_book, wang, yajima}

For an ensemble of noninteracting two-level systems having an energy separation $\hbar \omega_0$, the optical Bloch equations obtained using the density matrix formalism, the Heisenberg equation for time evolution and the relaxation time approximation, are a set of coupled equations connecting the time evolution of the off-diagonal and diagonal elements of the density matrix. These in turn are related to the polarization and population density respectively. Since the formalism is very standard, we will use the notation and methodology as is given in the book by Meier {\etal},~\cite{Meier-Thomas-Koch} where the perturbative solution of the optical Bloch equations under the two-beam delta-pulse excitation up to third order in polarization using rotating wave approximation is also calculated. The polarization terms calculated in this way (up to third order) contain various combinations of the wavevectors $\vec{k}_{pm}$ and $\vec{k}_{pr}$ arising from the combinations of the electric fields from the two incident laser beams. We pick up the terms having a phase factor $\exp(-i \vec{k}_{pr}\cdot\vec{r})$. These are responsible for generating the radiated electric field in the probe reflection direction. We have one first- and one third-order terms with this phase factor and the total polarization (up to third order) $\vec{P}_{total}(\vec{r},t,\Omega_{pr},\Omega_{pm})$ can be written as
\begin{equation} \label{ptotal}
\begin{split}
\vec{P}_{total}(\vec{r},t,\Omega_{pr},\Omega_{pm})=[\vec{a}_{1}(t,\Omega_{pr})\Theta(t)+\vec{a}_{2}(t,\Omega_{pr},\Omega_{pm})\Theta(\tau)\Theta(t-\tau)]e^{-\frac{t}{T_{2}}}\sin(\omega_0t-\vec{k}_{pr}\cdot\vec{r}),
\end{split}
\end{equation}
with
\begin{equation} \label{a1}
\vec{a}_{1}(t,\Omega_{pr})=\frac{2}{\hbar}\vec{\xi}^*_{pr}(t,\Omega_{pr})|\vec{\mu}_{vc}|^2,
\end{equation}
and
\begin{equation} \label{a2}
\vec{a}_{2}(t,\Omega_{pr},\Omega_{pm})=\frac{2}{\hbar^{3}}|\vec{\xi}_{pm}(t,\Omega_{pm})|^2\vec{\xi}^{*}_{pr}(t,\Omega_{pr})|\vec{\mu}_{vc}|^4e^{-\frac{\tau}{T_{1}}}.
\end{equation}
Here, $\mu_{vc}$ is the relevant dipole matrix element, $T_1$ and $T_2$, respectively, are the energy and phase relaxation times, and $\Theta$ is the Heaviside step function.

One can then take the Fourier transform of this polarization to get the expression for the third-order susceptibility which will be directly related to the dielectric constant measured in the absorption measurement. This is what is done in the treatment of Koch and coworkers.~\cite{Meier-Thomas-Koch} We shall consider coherent reflection instead. The case of reflectivity measurement is slightly different and one needs to explicitly consider the emitted radiation due to this polarization. As mentioned above, note that the position dependent part ($\vec{k}_{pr}\cdot\vec{r}$) of the phase of the oscillating polarization is responsible for the directionality of the emitted radiation.

%%%%%%%%%%%%%%%%%%%%%%%%%%%%%%%%%%%%%%%%
\begin{figure}[htb]
    \includegraphics[clip,width=8.0cm]{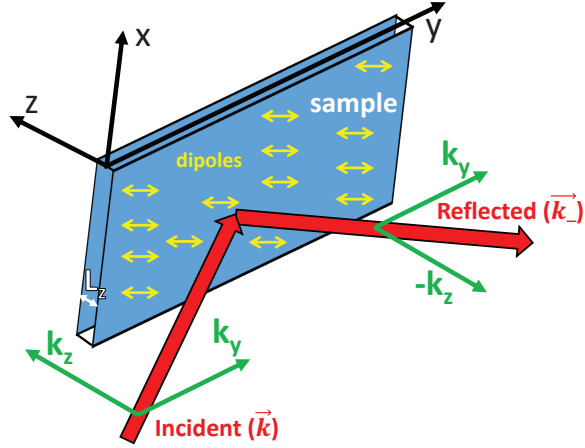}
    \caption{Schematic depicting a first principles view of reflection without invoking Snell's laws. The reflected beam is a result of the radiated electric field generated by the time dependent polarization current resulting from the electric dipoles. These dipoles are set oscillating by the incident beam.  Here the (quasi)-two-dimensional sample of thickness $L_z$ is considered to be of infinite extent in the $x$-$y$ plane and to be situated at $z=0$ plane. The $z$ component of the wave vector changes sign through reflection. Horizontal arrows depict oscillating dipoles responsible for the radiation emitted as the reflected wave. Free-induction-decay is simply the exponentially decaying free oscillations of these dipoles after they are excited by a short electric field pulse. The directionality of the reflected beam (angle of incidence equal to the angle of reflection) comes from the relative phase between the oscillating dipoles (constructive interference only in the reflection direction).}\label{ewald}
\end{figure}
%%%%%%%%%%%%%%%%%%%%%%%%%%%%%%%%%%%%%%%%%

In the absence of free charges, the current density $\vec{J}=d\vec{P}_{total}/dt$ generated in the sample due to the time-dependent polarization acts as the source for a vector potential of the radiated electric field. Note that the transition frequency $\omega_0$ between ground and excited states (here idealized as a two level system)  for a typical semiconductor sample falls in the optical frequency range ($\sim 10^{15}$~Hz). Thus the derivative of the $\sin(\omega_0t-\vec{k}_{pr}\cdot\vec{r})$ factor appearing in Eq.~\eqref{ptotal} will make a much larger contribution to the current density, compared to the other factors (the laser pulse width, the dipole oscillators' dephasing time or that of the mechanical chopper). Thus the current density may approximately be written as
\begin{equation}
%\begin{split}
\vec{J}(\vec{r},t,\Omega_{pr},\Omega_{pm})\cong \Re\left[\omega_0[\vec{a}_{1}(t,\Omega_{pr})\Theta(t)+ \vec{a}_{2}(t,\Omega_{pr},\Omega_{pm})\Theta(\tau)\Theta(t-\tau)]e^{-\Gamma t} e^{i({\omega_0t-\vec{k}_{pr}\cdot\vec{r}})}\right].
%\end{split}
\end{equation}
Here $\Gamma=T_2^{-1}$ and $\Re$ denotes the real part of the expression. This current density is the source of the radiated field in the inhomogeneous wave equation (Helmholtz equation)~\cite{Orfanidis}
\begin{equation}  \label{vecpot}
\vec{A}(\vec{r},t,\Omega_{pr},\Omega_{pm})=\int_{V} d^{3} r^{\prime}  \mu_{0}  \vec{J}(\vec{r^{\prime}},t)  G(|\vec{r}-\vec{r}^{\prime}|),
\end{equation}
where $\vec{A}(\vec{r},t,\Omega_{pr},\Omega_{pm})$ is the vector potential, $\mu_{0}$ is the permeability of the vacuum and
\begin{equation}
G(|\vec{r}-\vec{r}^{\prime}|)=\frac{e^{-i\vec{k}\cdot(\vec{r}-\vec{r}^{\prime})}}{4\pi |\vec{r}-\vec{r}^{\prime}|}
\end{equation}
is the Green's function for the Helmholtz equation, \textit{viz.},
\begin{equation}
\nabla^2 G(|\vec{r}-\vec{r}^{\prime}|)+k^2 G(|\vec{r}-\vec{r}^{\prime}|) = -\delta^{3}(|\vec{r}-\vec{r}\prime|).
\end{equation}
Here the spatial coordinate carrying $\vec{r}^\prime$ denotes the plane of the sample. Taking the incident wavevector to be $\vec{k}=k_{x}\hat{x}+k_{y}\hat{y}+k_{z}\hat{z}$ and assuming that the sample spans the entire $x$-$y$ plane at $z=0$ and it has negligible thickness ($L_{z}\to 0$) along the $z$ direction (Fig. 2), we can recast the integral of Eq.~\eqref{vecpot} as
\begin{equation}
\vec{A}(\vec{r},t,\Omega_{pr},\Omega_{pm})=\Re\left[\lim_{L_{z}\to 0} \int^{+\infty}_{-\infty}dx^{\prime}\int^{+\infty}_{-\infty} dy^{\prime} \int^{L_{z}}_{0} dz^{\prime} \mu_{0}  \vec{J}(\vec{r^{\prime}},t)  G(|\vec{r}-\vec{r}^{\prime}|)\right].
\end{equation}
This is readily integrated to yield~\cite{Orfanidis}
\begin{equation}
\begin{split}
\vec{A}(\vec{r},t,\Omega_{pr},\Omega_{pm})= \Re\left[\frac{\mu_{0}}{4\pi} e^{i\omega_0t}\omega_0[\vec{a}_{1}(t,\Omega_{pr})\Theta(t)+ \vec{a}_{2}(t,\Omega_{pr},\Omega_{pm})\Theta(\tau)\Theta(t-\tau)]e^{-\Gamma t}\right.\\ \left.\frac{e^{-i(k_{x}x+k_{y}y-k_{z}z)}}{2ik_{z}} \lim_{L_{z}\to 0} \int^{L_{z}}_{0}  dz^{\prime}   e^{i(k_{z}-k'_{z})z'}\right],
\end{split}
\end{equation}
where $k_{x}\hat{x}+k_{y}\hat{y}-k_{z}\hat{z}$ represents the reflected wave vector $\vec{k}_{ref}$. Proceeding with the integration along $z$ direction, we have
\begin{equation}
\begin{split}
\vec{A}(\vec{r},t,\Omega_{pr},\Omega_{pm})= \Re\left[\frac{\mu_{0}}{4\pi} e^{i\omega_0t}\omega_0[\vec{a}_{1}(t,\Omega_{pr})\Theta(t)+ \vec{a}_{2}(t,\Omega_{pr},\Omega_{pm})\Theta(\tau)\Theta(t-\tau)]e^{-\Gamma t}\right.\\ \left.\frac{e^{-i(k_{x}x+k_{y}y-k_{z}z)}}{2ik_{z}} \lim_{L_{z}\to 0} \frac{(e^{i(k_{z}-k'_{z}) L_{z}}-1)}{{i(k_{z}-k'_{z})}}\right].
\end{split}
\end{equation}
Note that $\lim_{L_{z}\to 0} [\{e^{i(k_{z}-k'_{z}) L_{z}}-1\}/i(k_{z}-k'_{z})]$ simply gives a factor of $L_z$.

Calculation of the reflected field in the probe direction is now straightforward~\cite{Orfanidis,lai} using the relation $\vec{E}_{rad}=-{\partial \vec{A}\over \partial t}-\vec{\nabla}\Phi$. As there are no free charges in the medium, the charge density $\rho=0$ and the term containing the scalar potential $\Phi$ is zero. Thus the radiated electric field in the probe reflection direction is
\begin{equation} \label{erad}
\begin{split}
\vec{E}_{rad}(t,\tau,\Omega_{pr},\Omega_{pm}, \vec{r})\simeq{\omega^{2}_{0}L_{z}\mu_{0}\over 4\pi}[\vec{a}_{1}(t,\Omega_{pr})\Theta(t)+\vec{a}_{2}(t,\Omega_{pm})\Theta(\tau)\Theta(t-\tau)] \\
e^{-\Gamma t}\frac{\cos[\omega_{0}t+\phi(\vec{r})]}{2k_{z}},
\end{split}
\end{equation}
where $k_{z}$ is the $z$-component of the wavevector. Again (due to the separation of time scales) we have only considered the most important term containing optical frequency $\omega_0$ while taking the time derivative of $\vec{A}$. The phase $\phi (\vec{r})=(\pi/2-\vec{k}_{ref}\cdot\vec{r})$. Note that the polarization in our calculation follows the conventional definition as being the induced dipole moment per unit volume. This is why the sample width $L_z$ explicitly appears the above expression. One may alternatively define a sheet polarization density without explicitly displaying $L_z$ as has been done, {\eg}  in Ref.~\onlinecite{shen_SHG} in the context of sum frequency generation using dipole radiation formula~\cite{jackson_electrodynamics_book} and where it was also shown that the radiated electric field is proportional to the induced polarization. Note that the above calculation is essentially an illustration of the idea behind the celebrated Ewald-Oseen extinction theorem~\cite{Wolf,Vincent,Berman} in the modern context of free-induction-decay at optical frequencies.

Coming back to the experiment, the perturbed-free-induction-decay is measured by the pump-probe differential reflectivity experiment at negative delay in which the probe pulse precedes the pump pulse by the delay time $\tau$. Since the reflected beam is passed through a monochromator which selects out one particular Fourier component (frequency) of the electric field, we express the reflected field (Eq.~\ref{erad}) as an integral over its Fourier modes:  
\begin{equation}
{\mathcal{\vec{E}}}_{rad}(\omega,\tau,\Omega_{pr},\Omega_{pm})=\frac{1}{\sqrt{2\pi}}\int^{+\infty}_{-\infty}{\vec{E}_{rad}(t,\tau,\Omega_{pr},\Omega_{pm}) e^{-i\omega t}dt}.
\end{equation}
Writing the cosine in the reflected filed as a sum of exponentials, we perform the integration and obtain
\begin{equation}
\begin{split}
\vec{\mathcal{E}}_{rad}(\omega,\tau,\Omega_{pr},\Omega_{pm})=\frac{\omega^{2}_{0}L_{z}\mu_{0}}{16\pi\sqrt{2\pi}k_{z}}\frac{[\vec{a}_{1}(t,\Omega_{pr})+\vec{a}_{2}(t,\Omega_{pr},\Omega_{pm})e^{-\Gamma \tau}e^{i(\omega_{0}-\omega)\tau}]}{\Gamma-i(\omega_{0}-\omega)}\\ + \frac{[\vec{a}_{1}(t,\Omega_{pr})+\vec{a}_{2}(t,\Omega_{pr},\Omega_{pm})e^{-\Gamma \tau}e^{i(\omega_{0}+\omega)\tau}]}{\Gamma+i(\omega_{0}+\omega)}.
\end{split}
\end{equation}
We may now neglect the terms having $\Gamma+i(\omega_{0}+\omega)$ in their denominator with respect to the terms having $\Gamma-i(\omega_{0}-\omega)$ in their denominator. Such an approximation can be appreciated when one considers that the first denominator is about three orders of magnitude larger than the second one for the case $\omega \cong \omega_{0}$. Then we have
\begin{equation}
\begin{split}
\vec{\mathcal{E}}_{rad}(\omega,\tau,\Omega_{pr},\Omega_{pm})=\frac{\omega^{2}_{0}L_{z}\mu_{0}}{16\pi\sqrt{2\pi}k_{z}}\frac{[\vec{a}_{1}(t,\Omega_{pr})+\vec{a}_{2}(t,\Omega_{pr},\Omega_{pm})e^{-\Gamma \tau}e^{i(\omega_{0}-\omega)\tau}]}{\Gamma-i(\omega_{0}-\omega)}.
\end{split}
\end{equation}
The action of the monochromator may be approximated by inserting a $\delta$-function frequency filter around the selected frequency $\omega_{m}$ while taking the inverse Fourier transform, i.e.,
\begin{equation}
\vec{E}^{Mono}_{rad}(t,\tau,\omega_{m},\Omega_{pr},\Omega_{pm})=\frac{1}{\sqrt{2\pi}}\int^{+\infty}_{-\infty}
\vec{\mathcal{E}}_{rad}(\omega,\tau,\Omega_{pr},\Omega_{pm})e^{i\omega t}\delta(\omega-\omega_{m})d\omega.
\end{equation}
This gives
\begin{equation}
\begin{split}
\vec{E}^{Mono}_{rad}(t,\tau,\Omega_{pr},\Omega_{pm})=\frac{\omega^{2}_{0}L_{z}\mu_{0}}{32\pi^2 k_{z}}\frac{[\vec{a}_{1}(t,\Omega_{pr})+\vec{a}_{2}(t,\Omega_{pr},\Omega_{pm})e^{-\Gamma \tau}e^{i(\omega_{0}-\omega{m})\tau}]}{\Gamma-i(\omega_{0}-\omega)}e^{i\omega_{m}t}.
\end{split}
\end{equation}
The detector measures the intensity integrating over a period of time ($\sim$ ns-$\mu$s) that is much larger than time period of oscillation of the electric field and much less than the chopping timescale (which is akin to a coarse graining). So for a detection time scale $1/\omega<<T<<1/\Omega_{pr/pm}$ the time integrated signal can be calculated as
\begin{equation}
S(t,\tau,\omega_{m},\Omega_{pr},\Omega_{pm})=\frac{1}{T}\int_{0}^{T}|\vec{E}^{Mono}_{rad}(t,\tau,\omega_{m},\Omega_{pr},\Omega_{pm})|^2dt,
\end{equation}
which contains the products of Eq.~\eqref{a1} and Eq.~\eqref{a2} as well as their individual squares which will have sine and cosine terms containing various combinations of the chopping frequencies $\Omega_{pr}$ and $\Omega_{pm}$. Note that these are low-frequency modulations created by the optical chopper, over and above the high optical frequency oscillation of the electric field. For phase-sensitive detection, a lock-in amplifier multiplies the signal received from the detector with a reference signal given to the lock-in amplifier [for double modulation the multiplicative term is $\cos(\Omega_{pr}+\Omega_{pm})t$].~\cite{sandip_double-modulataion}  After passing through a low pass filter in the lock-in amplifier only the components having $(\Omega_{pr}+\Omega_{pm})$ frequency will survive and all other frequency terms will be rejected. Finally the measured signal has the form
\begin{equation}
\begin{split}
S(t,\tau,\omega_{m},\Omega_{pr},\Omega_{pm}) =-{1\over 2}\left[\frac{\omega^{2}_{0}L_{z}\mu_{0}}{ 32 \pi^{3} k_{z}}\right]^2\frac{\cos(\theta_{1}+\theta_{2})}{(\Gamma^2+(\omega_{0}-\omega_{m})^2)} \bigg[ \frac{8}{\hbar^{4}} |\mu_{vc}|^{6}|\vec{\zeta}_{pr}|^2 |\vec{\zeta}_{pm}|^2 e^{-\tau/T_{1}} e^{-\Gamma \tau}\\ \cos(\omega_{0}-\omega_{m})\tau  + \frac{4}{\hbar^{6}}e^{-2\tau / T_{1}}|\mu_{vc}|^{8} e^{-2\Gamma \tau } |\vec{\zeta}_{pr}|^2 |\vec{\zeta}_{pm}|^4 \bigg].
\end{split}
\end{equation}

The measured signal has two components, the first one is the signature of the cross term of the kind $\vec{a}^{*}_{1}(t,\Omega_{pr})\vec{a}_{2}(t,\Omega_{pr},\Omega_{pm})$ or its complex conjugate, while the second term is just the contribution from $|\vec{a}_{2}(t,\Omega_{pr},\Omega_{pm})|^{2} e^{-2 \Gamma \tau}$. The second term being two orders higher in the pump fluence is much smaller in comparison with the first term and for a fixed delay provides only a constant background upon which the frequency dependence are observed. Therefore it is neglected with respect to the first term and we have the expression (up to a scale factor denoting the detector efficiency) for the dc voltage $V_\textrm{lockin}$ finally displayed by the lock-in amplifier for a given $\tau$ and $\omega_m$
\begin{equation}
V_\textrm{lockin} =-\frac{\omega^{4}_{0}L_{z}^2\mu_{0}^2}{ 256 \pi^{6} k_{z}^2\hbar^{4}}\frac{\cos(\theta_{1}+\theta_{2})}{(\Gamma^2+(\omega_{0}-\omega_{m})^2)} |\mu_{vc}|^{6}|\vec{\zeta}_{pr}|^2 |\vec{\zeta}_{pm}|^2 e^{-\tau/T_{1}} e^{-\Gamma \tau}\cos(\omega_{0}-\omega_{m})\tau.
\end{equation}

The theoretically calculated signal has the following important features:
(i) the signal is nonzero at negative delay and it is exponentially decaying with $\tau$ with a time constant $\Gamma = T_2^{-1}$ when monochromator is tuned at the transition frequency, i.e.,  $\omega_{m} = \omega_{0}$, (ii) if the signal is detected at a frequency other than $\omega_{0}$, an oscillatory feature is superimposed on the exponentially decaying part of the signal as a function of delay, where the oscillation period depends on the detuning $\omega_{m}-\omega_{0}$,
(iii) for a fixed $\tau$, the signal oscillates as a function of $\omega_{m}$, the detection frequency set at the monochromator {\ie} coherent spectral oscillations are predicted, and (iv) $\int S(\tau,\omega_{m})d\omega_{m}\approx 0$ due to the oscillatory cosine term in the signal. This implies that without the monochromator the signal is zero which has been reported earlier.~\cite{Guenther,Joffre} It is worth noting that while the inverse Lorentzian shape of the signal and coherent oscillations have been reported for negative delay for differential absorption,~\cite{Meier-Thomas-Koch} our form for probe reflected signal consists of a single term and is quite different. For zero delay the coherent oscillations vanish and we get a purely negative bleaching.

%%%%%%%%%%%%%%%%%%%%%%%%%%%%%%%%%%%%%%%%
\begin{figure}[htb]
    \includegraphics[clip,width=8.0cm]{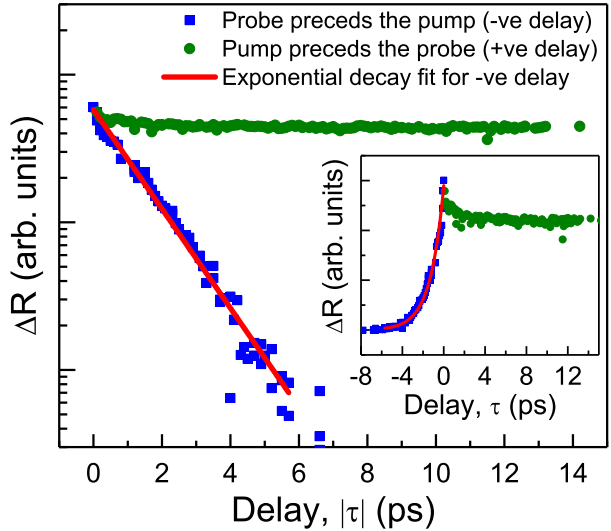}
    \caption{The time evolution of the PPDR signal at the hh-x peak for $17.5$ nm QW sample measured at $4$~K. Note that only the magnitude of delay is plotted along x-axis. Solid line is an exponential fit to the data at negative delay region  giving the dephasing time to be $1.5$ ps. (inset) The original data in linear scale for both positive and negative delays, as well as the exponential fit at negative delays.}\label{ple}
\end{figure}
%%%%%%%%%%%%%%%%%%%%%%%%%%%%%%%%%%%%%%%%%

%%%%%%%%%%%%%%%%%%%%%%%%%%%%%%%%%%%%%%%%
\begin{figure}[htb]
	\includegraphics[clip,width=8.0cm]{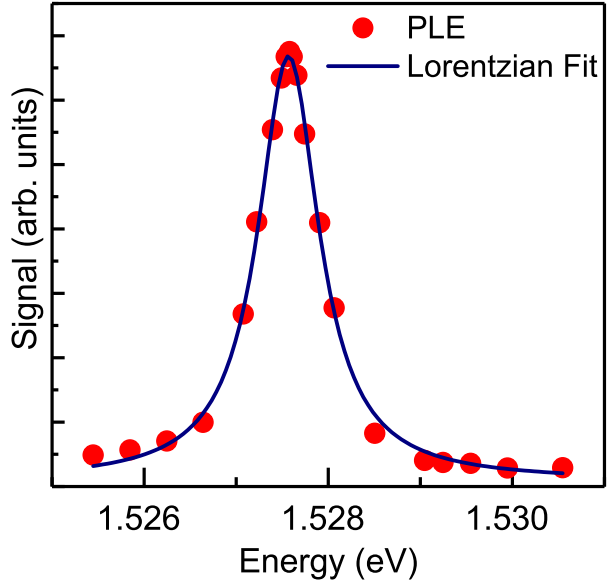}
	\caption{Continuous wave PLE spectrum for the $17.5$ nm QW sample is plotted around the hh-x peak. The solid line is a Lorentzian fit which gives the line width of the spectrum corresponding to the dephasing time ($T_{2}$) of $1.68$ ps which is in excellent agreement with the value estimated in Fig.~3.}\label{ple2}
\end{figure}
%%%%%%%%%%%%%%%%%%%%%%%%%%%%%%%%%%%%%%%%%

\section{\label{sec:level3}Experimental Results}Let us now compare this theory with the experiment performed at low temperature (4~K) on a canonical sample, high quality GaAs QWs. The experimental arrangement is already described in Fig. 1. The Fourier transform broadened ($\approx 10$ meV)  degenerate pump and probe pulsed laser beams ($\approx 100$ fs) were  frequency tuned to the heavy-hole exciton (hh-x) resonance ($\approx 1.5275$ eV) in this multi-quantum well sample having the well-width of $17.5$~nm. Figure 3 shows the delay dependence of the PPDR signal measured after it was passed through the monochromator with the bandpass of the monochromator set at the hh-x resonance frequency where the signal at zero delay was of maximum amplitude. The PPDR signal measured at the hh-x peak in the negative delay region (where the probe precedes the pump) survives up to about 6~ps. Note the logarithmic scale along $y$-axis in Fig. 3 represents the magnitude of the signal measured by the lock-in amplifier. The straight line behaviour on the semilogarithmic scale is indicative of the exponentially decrease, as predicted by Eq. (23). This exponentially decaying signal is the signature of the free-induction-decay. To extract the dephasing time $(T_{2} = 1/\Gamma$), the PPDR signal in the negative delay is fitted with a single exponential function as shown in Fig.~\ref{ple}. The fitting results to $T_{2} = 1.5$~ps. Note that $T_1 >> T_2$ in Eq. (23) and hence the effect of $T_1$, the radiative decay time, is safely ignored.

Figure 4 shows the photoluminescence excitation (PLE) spectrum (which is equivalent to measuring the absorption spectrum~\cite{Pelant-Valenta}) at the hh-x peak. The Lorentzian fit to the PLE spectrum gives the full-width-at-half-maximum (FWHM) equivalent to the dephasing time $T_{2} \: (= 2\hbar /\text{FWHM})$~\cite{Honold} of $1.68$ ps which closely matches with the dephasing time estimated from th exponential fit in Fig. 3. This supports the validity of our PPDR measurements.

Finally, coherent spectral oscillations~\cite{Fluegel, Meier-Thomas-Koch}  can be seen in Fig.~5 where the spectrally-resolved PPDR signal around the hh-x resonance is plotted for the negative delay values fixed at $0,\, 2,\, 4,\, $ and $6$ picoseconds. The theoretically calculated spectra using Eq.~(23) for $\hbar \omega_0 = 1.5275$~eV and $\hbar \Gamma = 0.4$~meV are shown in the accompanying figures at corresponding delays. They qualitatively match the experimental spectra as they reproduce the coherent spectral oscillation and the exponential decrease of the signal near hh-x peak with increasing delay. Deviations on the lower energy side may be explained by the presence of at least one additional lower energy bound exciton resonance,  which is not included in the theoretical calculation. Note that the coherent spectral oscillations vanish with approaching zero delay. The oscillation frequency increases with increasing negative delay due to the $\cos[(\omega_0 - \omega_m)\tau]$ dependence.

%%%%%%%%%%%%%%%%%%%%%%%%%%%%%%%%%%%%%%%%
\begin{figure}[htb]
    \includegraphics[clip,width=10.0cm]{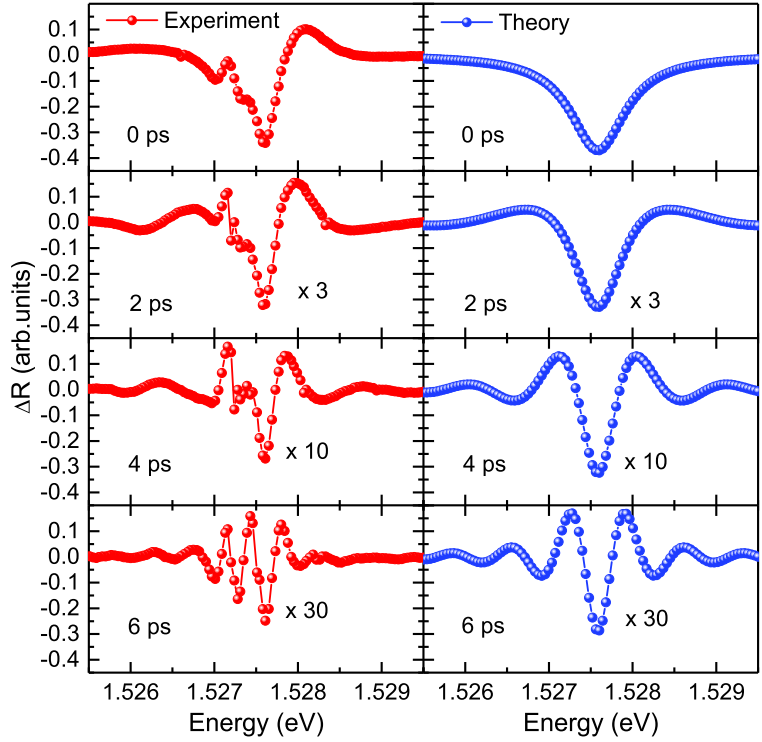}
    \caption{(Left column) Spectrally-resolved pump-probe reflectivity signal for different negative delays for the 17.5~nm QW sample at $4$~K. The pump and the probe powers were 10~mW and 0.5~mW respectively. (Right column) Theoretically calculated spectra using Eq.~(23) for $\hbar \omega_0 = 1.5275$~eV and $\hbar \Gamma = 0.4$~meV at the same negative delay values as used in the left column. Note that identical scales are set along the $x$- and $y$-axes for the plots of the experimental and theoretical spectra at different delays. The signal is multiplied by a suitable multiplicative factor to fill the real estate of the plot. While the experimental spectra suggest the presence of a weaker (defect related) second resonance at a slightly lower energy, the theoretical fits [Eq. (23)] used only a single resonance to minimize the number of free adjustable parameters.}\label{cohosc}
\end{figure}
%%%%%%%%%%%%%%%%%%%%%%%%%%%%%%%%%%%%%%%%%

\section{Summary and Outlook}
Given how non-intuitive the interpretation of the negative-delay pump probe differential reflectivity signal is, we felt that it will be useful to explicitly work out a simplified closed-form expression for the actual reflectivity signal one would measure in the experiment, mathematically accounting for all the optical and electronic components. The analysis clearly brought out the following specific features of the third order perturbed free induction decay signal: (i) the signal is nearly zero without the monochromator, (ii) it would scale proportional to both the pump and the probe intensities, (iii) the signal measured at resonance energy decays exponentially with the (negative) delay with the value of the decay constant being the dephasing time, (iv) the signal for a fixed value of negative delay shows coherent spectral oscillations as a function of energy. The predictions were explicitly tested in a low temperature PPDR experiment on GaAs quantum well sample, where coherent oscillations were observed to survive up to a negative delay of 6 ps between the probe and the pump beam. Our experimental data has a reasonable match with the theoretically calculated output signal. The presence of lower energy bound states make the signal slightly deviate from the theoretical one. Perhaps the most noteworthy is the close agreement of the experimentally obtained dephasing time from the perturbed-free-induction-decay  with the inverse linewidth of the absorption profile of the damped driven oscillator (inferred from the continuous wave PLE spectrum linewidth). Given the generality and simplicity of our analysis, we hope that the PPDR experiments will become more popular and will be further used to study other systems having plasmonic and other quasiparticle resonances that may arise out of strong electron correlation.


\begin{thebibliography}{99}
\bibitem{fnote1} In the standard two-beam degenerate four-wave-mixing experiments, ultrashort pulses from two laser-beams having wavevectors $\vec{k}_1$  and $\vec{k}_2$ are cofocused on to the sample and the four-wave-mixing signal arising from the third order nonlinear optical effect is detected in the background-free direction given by $2\vec{k}_2-\vec{k}_1$. Because there is no linear optical signal propagating in the $2\vec{k}_2-\vec{k}_1$ direction to guide, identifying four-wave-mixing signal propagation direction is experimentally a challenging task.
\end{thebibliography}
\end{document}